\documentclass[twocolumn,showpacs,preprintnumbers,superscriptaddress,amsmath,amssymb,floatfix,aps]{revtex4}

\usepackage[pdftex]{graphicx}
\usepackage{dcolumn}
\usepackage{bm}
\begin{document}

\title{Graphene-on-silicon near-field thermophotovoltaic cell}

\author{V. B. Svetovoy}
\affiliation{MESA$^+$ Institute for Nanotechnology, University of
Twente, PO 217, 7500 AE Enschede, The Netherlands}
\affiliation{Institute of Physics and Technology, Yaroslavl Branch,
Russian Academy of Sciences, 150007, Yaroslavl, Russia}
\author{G. Palasantzas}
\affiliation{Zernike Institute for Advanced Materials, University of
Groningen - Nijenborgh 4, 9747 AG Groningen, The Netherlands}

\date{\today}

\begin{abstract}
A graphene layer on top of a dielectric can dramatically influence
ability of the material to radiative heat transfer. This property of
graphene is used to improve the performance and reduce costs of
near-field thermophotovoltaic cells. Instead of low bandgap
semiconductors it is proposed to use graphene-on-silicon Schottky
photovoltaic cells. One layer of graphene absorbs around 90\%  of
incoming radiation and increases the heat transfer. This is due to
excitation of plasmons in graphene, which are automatically tuned in
resonance with the emitted light in the mid infrared range. The
absorbed radiation excites electron-hole pairs in graphene, which are
separated by the surface field induced by the Schottky barrier. For a
quasi-monochromatic source the generated power is one order of
magnitude larger and efficiency is on the same level as for
semiconductor photovoltaic cells.

\end{abstract}

\maketitle

\section{Introduction}

Thermophotovoltaic (TPV) generators convert radiation emitted by a
heated body into electricity \cite{Cou01}. In these devices a hot
emitter radiates the electromagnetic energy that is absorbed by a
cold photovoltaic (PV) cell (collector). In the solar TPV generators
\cite{Len14} solar light is absorbed and then re-emitted as thermal
radiation in a spectrally selective way. Any other heat source also
can be used for transformation into electricity: wasted industrial
heat, the heat from car engines, computer chips etc.
\cite{Bas09,Bos12}. TPV systems are expected to be quiet, modular,
safe, low-maintenance, and pollution-free \cite{Cou01}.

The main challenge is to increase both efficiency and output electric
power of the devices. The power is restricted by the emissivity of
the black body. This restriction, however, is not applicable when the
bodies are separated by a distance much smaller than the thermal
wavelength \cite{Pol71}. In this near-field range the radiative heat
transfer (RHT) can be increased by orders of magnitude
\cite{Pen99,Par02,Mul02}. Significant enhancement of the RHT in the
near-field was demonstrated experimentally
\cite{Har69,Kit05,Hu_08,Nar08,She09,Rou09}. The highest efficiency is
reached when the emitter is a nearly monochromatic source of
radiation with the photon energy slightly larger than the band gap of
the PV cell \cite{Lar06}. On the other hand, the strongest
enhancement is realized as the resonance energy exchange when both
the emitter and collector support surface modes such as plasmon or
phonon-polaritons matching each other \cite{Jou05,Vol07}. The
absorption of light by semiconductor PV cells does not have resonance
character and cannot support high RHT. Moreover, TPV generators have
to use narrow band semiconductors, which are much more expensive than
silicon. In this paper we explain how graphene could help to resolve
these issues.

Graphene can add new functionalities to materials that do not have
surface modes in the mid infrared (IR) range \cite{Sve12}. This is
because graphene supports plasmons with the frequency that is varied
with the wave number, coupling constant, and Fermi level
\cite{Jab09}. The plasmon frequency in graphene is automatically
tuned with the frequency of the surface mode in the opposing body
resulting in a significant increase of the RHT \cite{Sve12,Ili12b}.
This prediction was also confirmed experimentally \cite{Zwo12}.

Here we propose to use  graphene-on-silicon (g/Si) Schottky
photodiode as a PV cell. In such a generator the emitted radiation is
resonantly absorbed in graphene where it excites electrons able to
overcome the Schottky barrier, there is no p-n junction and related
optical losses in the low price Si substrate, there is no problem to
couple the evanescent radiation to electrons in graphene, and the
device has a simple structure. The silicon substrate is transparent
in the wavelength range $\lambda=1.2-8.0\; \mu\textrm{m}$ providing
the optimal conditions for the RHT \cite{Sve12}.

The g/Si Schottky photodiodes were already applied for solar cells
\cite{Li_10,Ye_12,Mia12}, where the visible light generated
electron-hole pairs in Si separated by the surface field. The
photodiode presented in \cite{Wan13} absorbs mid IR light in the
graphene layer and the Schottky barrier separates electrons and
holes. The responsivity of the device was estimated as at least 0.13
A/W.

Application of graphene to enhance performance of the near-field TPV
generators was already discussed. A free-standing graphene emitter
was considered in combination with the low bandgap (0.17 eV) InSb
photovoltaic cell \cite{Ili12a}. For this device there is no
resonance energy exchange between graphene and the semiconductor. A
PV cell consisting of InSb substrate covered with graphene was
discussed in \cite{Mes13}, where hexagonal Boron Nitride (hBN) was
used as the emitter. In this device the RHT is enhanced due to the
resonance tuning but most of the photons are absorbed in graphene
without generation of photoelectrons.

\section{Heat transfer}

\begin{figure}[ptb]
\begin{center}
\includegraphics[width=86mm]{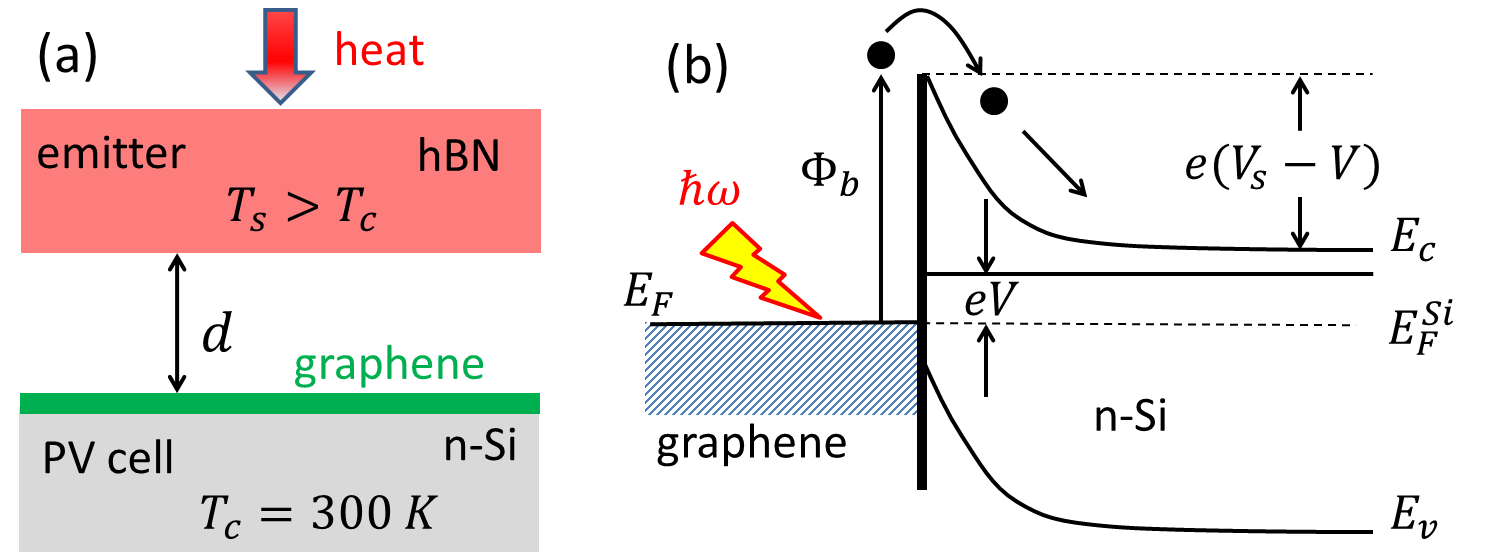}
\vspace{-0.3cm} \caption{(Color online) (a) Schematic representation of the
TPV element. (b) Energy diagram for the Schottky graphene-on-Si diode working
as an infrared PV cell. }
\label{fig1}
\vspace{-0.9cm}
\end{center}
\end{figure}

We consider graphene-on-silicon Schottky junction in combination with
the hBN emitter that gives quasi-monochromatic radiation at 0.195 eV
(see Fig. \ref{fig1}(a)). This specific emitter is not the point of
interest as long as the photoelectrons can overcome the Schottky
barrier. Moreover, a wider range of emitted frequencies is
preferable. It is assumed that the bottom of the PV cell is kept at
room temperature $T_c=300$ K but the emitter temperature $T_s$ can
vary.

The energy scheme of the photodiode is shown in Fig. \ref{fig1}(b).
Silicon of n-type is shown but p-type also can be used. Infrared
radiation from the emitter is absorbed in graphene providing hot
electrons that are able to overcome the Schottky barrier
$\Phi_b=E_F-\chi_{Si}$, where $E_F$ is the Fermi level in graphene
and $\chi_{Si}$ is the electron affinity in silicon. The surface
barrier from the Si side is $V_s$ and the forward bias applied to the
diode is $V$. The edges of the conduction and valence bands are $E_c$
and $E_v$, respectively, the Fermi level in Si is $E_F^{Si}$.

In contrast with the ordinary PV cells Si is transparent for the
emitted radiation. The Fermi level in pristine graphene is
$E_F^0=4.56$ eV \cite{Yan12b} but it can be adjusted by chemical
doping of graphene in a wide range \cite{Jeo10,Yi_11,Shi10}. The
barrier height has to be $\Phi_b=0.19$ eV or lower if hBN emitter is
used. This height can be reached for n-doping in graphene with the
relative Fermi level ${\cal E}_F=E_F-E_F^0=0.32$ eV counted from the
Dirac point. The band bending in Si is defined by the surface
potential $eV_s =\Phi_b-(E_F^{Si}-E_c)$.

The photocurrent is generated by photons with the energy $\hbar\omega
> \Phi_b$ absorbed in the graphene layer. The radiation of emitter
absorbed in graphene $R_g$ can be calculated as
\begin{equation}\label{Phi_g}
    R_g(T_s,d) = R^-(T_s,d) - R^+(T_s,d).
\end{equation}
Here $R^-(T_s,d)$ is the heat flux from the emitter to the collector
taken in the gap just above the graphene layer. This flux is
calculated as the $z$-component of the averaged Pointing vector
induced by the fluctuations in the emitter. The flux $R^+$ is
calculated in a similar way but just below the graphene layer. These
fluxes can be presented in the form
\begin{equation}\label{R_def}
    R^{\pm}(T_s,d) = \int_0^{\infty}\frac{d\omega }{2\pi}\hbar\omega
    n_B(\omega,T_s)\Pi^{\pm}(\omega,d),
\end{equation}
where $n_B(\omega,T)=(e^{\hbar\omega/T}-1)^{-1}$ is the Bose factor
(the Boltzmann constant here is $k_B=1$). The evanescent spectral
function $\Pi^-(\omega,d)$ is
\begin{equation}\label{Pi_def}
    \Pi^-(\omega,d)=\int\frac{d^2q}{(2\pi)^2}\frac{4\:
    \textrm{Im}r_c\:\textrm{Im}r_s e^{-2qd}}
    {\left|1-r_s r_c e^{-2qd}\right|^2}.
\end{equation}
Here $r_s$ and $r_c$ are the reflection coefficients of the source
and collector, respectively, which are functions of $\omega$ and the
wave vector $\textbf{q}$ parallel to the plates. The result for
$\Pi^-(\omega,d)$ is well known \cite{Pol71,Mul02,Jou05,Vol07} but
the expression for $\Pi^+(\omega,d)$ has to be calculated. It can be
done using the standard approach of fluctuational electrodynamics.
Skipping the details, the final expression for the spectral function
$\Pi_g=\Pi^--\Pi^+$ is
\begin{equation}\label{Pi_g_def}
    \Pi_g(\omega,d)=\int\frac{d^2q}{(2\pi)^2}\frac{4\kappa\:
    \textrm{Im}\varepsilon_g\:\textrm{Im}r_s
    \left|1-r_c\right|^{2} e^{-2qd}}
    {\left|1-r_s r_c e^{-2qd}\right|^2},
\end{equation}
where $\varepsilon_g(\omega,q)$ is the dielectric function of
graphene and $\kappa$ is the average dielectric constant of the media
above and below the graphene layer.

Graphene is responsible for the energy exchange in the system.
Without graphene on top of silicon the heat transfer is negligible
because Si is transparent for the emitteed radiation.  It is known
that the effect of graphene can be evaluated with a good precision
($\sim \alpha=e^2/\hbar c$) in the non-retarded limit
$c\rightarrow\infty$ \cite{Gom09,Sve11,Sve12}, where only p-polarized
evanescent fluctuations contribute to the momentum or energy exchange
between parallel plates separated by a submicrometer gap. Therefore,
in Eqs. (\ref{R_def})-(\ref{Pi_g_def}) it is sufficient to take into
account only the contribution of p-polarized evanescent waves.

In order to calculate the total energy exchange between bodies we
have to include the heat flux going to the emitter and originating
from fluctuations in the collector. Normally  it can be done by the
substitute in Eq. (\ref{R_def})
\begin{equation}\label{dir-rev}
    n_B(\omega,T_s)\rightarrow N(\omega,T_s,T_c)= n_B(\omega,T_s)-
    n_B(\omega,T_c)
\end{equation}
because in the thermal equilibrium $T_s\rightarrow T_c$ the total RHT
has to be zero \cite{Pol71}. However, the procedure is different if
the collector is used as a PV element. As for semiconductors
\cite{Wur82,Luq03} the potential difference $V$ results in
non-thermal photons that have non-zero chemical potential
$\mu_{ph}=eV$  but still can be described by the collector
temperature $T_c$. This is true for photons with the energy above the
Schottky barrier $\hbar\omega>\Phi_b$; the photons with smaller
energies have $\mu_{ph}=0$. In this case the thermal factor
$N(\omega,T_s,T_c)$ is defined as
\begin{equation}\label{dir-rev_non}
    N(\omega,T_s,T_c)= n_B(\omega,T_s)-
    n_B(\omega-\mu_{ph}/\hbar,T_c),
\end{equation}
where we have to understand $\mu_{ph}$ as a discontinuous function of
$\omega$
\begin{equation}\label{mu_def}
    \mu_{ph}(\omega)=\left\{\begin{array}{c}
                      0\ \ \ \hbar\omega<\Phi_b, \\
                      eV\ \ \ \hbar\omega>\Phi_b.
                    \end{array}
    \right.
\end{equation}

The final result for the radiative power $P_{rad}$ per unit area is
\begin{equation}\label{P_abs}
    P_{rad}(d,T_s,T_c) = \int_0^{\infty}\frac{d\omega }{2\pi}\hbar\omega
    N(\omega,T_s,T_c)\Pi^{-}(\omega,d).
\end{equation}
It is similar to the expression used in \cite{Ili12a,Mes13}. The
difference is that in our case the collector emits radiation with
$\hbar\omega< \Phi_b$ due to presence of graphene, while for a
semiconductor PV cell with the bandgap $E_{gap}$ the adopted
approximation is that the radiation with $\hbar\omega< E_{gap}$ is
not emitted. The radiative power absorbed in the graphene layer $P_g$
can be calculated from (\ref{P_abs}) with the substitute
$\Pi^-\rightarrow \Pi_g$.

\section{Generated power}

Let us assume first that each photon with the energy above $\Phi_b$
which is absorbed in graphene produces one electron in the conduction
band of Si. The actual responsivity of the photodiode will be
discussed later. In this case the photocurrent generated in the cell
\cite{Lar06} is
\begin{equation}\label{Iph}
    I_{ph}(V) = e\int_{\Phi_b/\hbar}^{\infty}\frac{d\omega }{2\pi}
    N(\omega,T_s,T_c)\Pi_{g}(\omega,d).
\end{equation}
It is proportional to the number of photons with the energy above
$\Phi_b$ absorbed in graphene. The generated electrical power and
efficiency of the cell are defined as
\begin{equation}\label{P_PV}
    P_{PV}=V\:I_{ph}(V),\ \ \ \eta=P_{PV}/P_{rad},
\end{equation}
where $P_{PV}$ is similar to that used in Refs. \cite{Ili12a,Mes13}.
Note that $\eta$ does not include the efficiency of heating of the
emitter and the efficiency of light transformation by the photodiode.
The electric power is zero for $V=0$ corresponding to the short
circuit and for $V=(1-T_c/T_s)\Phi_b/e$ corresponding to the open
circuit voltage. The maximal power is realized somewhere in between
these values.

The dielectric function of the emitter is described by the
Drude-Lorentz model
\begin{equation}\label{eps_hBN}
    \varepsilon_{hBN}(\omega)=\varepsilon_{\infty}\left(1+
    \frac{\omega_L^2-\omega_T^2}{\omega_T^2-\omega^2-i\Gamma\omega}\right)
\end{equation}
with the parameters of hBN from \cite{Nar03}
$\varepsilon_{\infty}=4.88$, $\omega_L=0.2$ eV, $\omega_T=0.17$ eV,
and $\Gamma=0.66\times 10^{-3}$ eV. The reflection coefficient of the
emitter (p-polarization, non-retarded)
$r_s=(\varepsilon_{hBN}-1)/(\varepsilon_{hBN}+1)$ has a surface
phonon-polariton resonance at 0.195 eV.

The reflection coefficient of the collector $r_c$ can be presented in
the form
\begin{equation}\label{r_c}
    r_c=\frac{\varepsilon_{Si}-1+2\kappa(\varepsilon_g-1)}
    {\varepsilon_{Si}+1+2\kappa(\varepsilon_g-1)},
\end{equation}
where $\varepsilon_{Si}\approx 11.9$ is practically a constant for Si
in mid IR and $\varepsilon_g(\omega,q)$ is the dielectric function of
graphene. In the limit of small relaxation frequency
$\gamma\rightarrow 0$ the latter can be presented in the form
\begin{equation}\label{eps_g}
    \varepsilon_g(\omega,q)=1+\frac{4\alpha_g {\cal E}_F}{\hbar v_F q}
    \left(1-\frac{\omega}{\sqrt{\omega^2-v_F^2q^2}}\right),
\end{equation}
where $\alpha_g=e^2/\kappa\hbar v_F$ is the coupling constant and
$v_F$ is the Fermi velocity in graphene. Equation (\ref{eps_g}) can
be applied in the range $q<2T/\hbar v_F$; finite $\gamma$ can be
accounted with the substitute $\omega\rightarrow\omega+i\gamma$ and
some not essential modification of $\varepsilon_g$ (see
\cite{Sve12,Ili12b} for details).

The spatial dispersion of $\varepsilon_g$ is an important property.
It results in the plasmon frequency that depends on $q$ even in the
non-retarded limit:
\begin{equation}\label{plasma}
    \hbar\omega_p(q)\approx \left(2\alpha_g\hbar v_F q {\cal E}_F\right)^{1/2}.
\end{equation}
This dependence means that for a body covered with graphene one can
always find a value of $q$ that gives the plasmon resonance matching
the surface mode in the opposite body. When the substrate
permittivity is large, graphene gives only small correction to $r_c$
and there will be no significant increase in RHT. For silicon it is
rather large but even a thin native oxide ($h\sim 1$ nm) on Si
influences the reflection coefficient. We take this oxide into
account in our calculations and use the graphene relaxation frequency
$\gamma=33$ meV ($5\times10^{13}$ rad/s).

\begin{figure*}[ptb]
\begin{center}
\includegraphics[width=120mm]{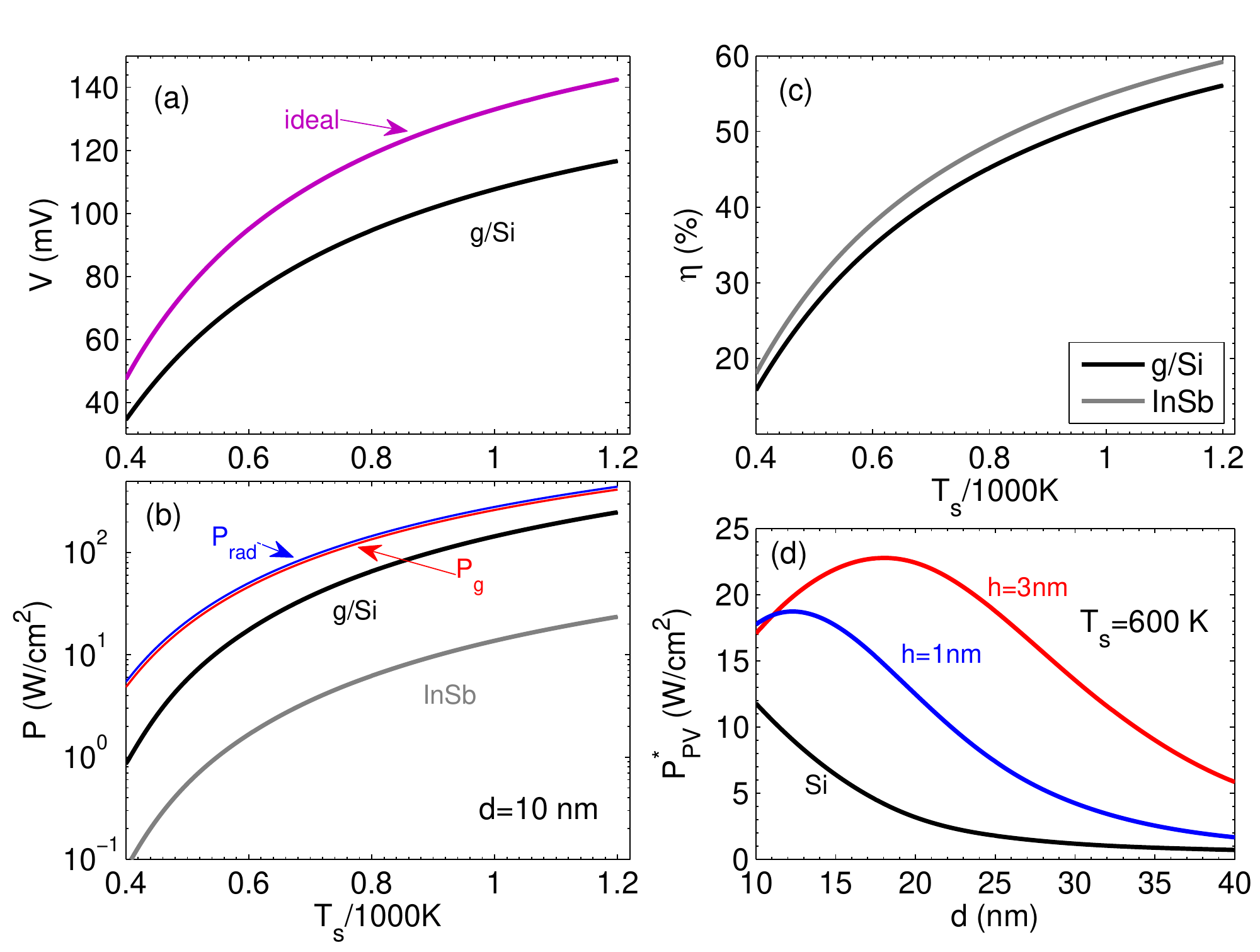}
\vspace{-0.5cm} \caption{(Color online). (a) The theoretical limit
on the operating voltage (ideal) and the voltage maximizing $P_{PV}$ for
g/Si cell as functions of $T_s$. (b) The radiative (blue) and absorbed
in graphene (red) powers as functions of temperature. The output power
for the g/Si cell and InSb cell are presented as black and gray curves,
respectively. (c) Efficiency of the TPV elements as a function
of temperature. (d) Dependence of the scaled output power on
the distance for different thicknesses of the SiO$_2$ layer on Si
including bare substrate.}
\label{fig2}
\vspace{-0.9cm}
\end{center}
\end{figure*}

The results are shown in Fig. \ref{fig2}. First, using Eq.
(\ref{P_PV}) the output power was maximized by varying the operating
voltage of the cell $V$. The theoretical limit of this voltage $V<
(1-T_c/T_s)\Phi_b/e$ and the value found from the maximization of
$P_{PV}$ at $d=10$ nm are shown in Fig. \ref{fig2}a as functions of
the emitter temperature. The optimal value of $V$ varies only
slightly with the distance. The net radiative power $P_{rad}$ and the
power absorbed in graphene $P_g$ are shown in Fig. \ref{fig2}b. The
curves are nearly coincide because about 90\% of incoming radiation
is absorbed in graphene. This exceptional phenomena is the result of
plasmon excitation in resonance with the radiation of the emitter.
The output power of the cell is also presented in Fig. \ref{fig2}b.
For comparison $P_{PV}$ for the InSb PV cell is shown too, which is
one order of magnitude smaller. The efficiency for g/Si cell shown in
Fig. \ref{fig2}c is somewhat smaller than that for the InSb cell. It
happens because 90\% of the light incoming on the g/Si cell generates
photocarriers, while it was assumed that 100\% of the light generates
carriers in the semiconductor cell. The latter, of course, is not
true because to contribute to the photocurrent the electron-hole
pairs has to reach the depletion layer before recombination
\cite{Par08,Fra11}. We can conclude that the TPV element with the
graphene-on-Si Schottky PV cell can outperform semiconductor PV cells
if the responsivity of the Schottky photodiode will be comparable
with that of low bandgap semiconductors.

The dependence of the scaled output powers $P_{PV}^* =
P_{PV}\times(d/10\textrm{nm})^2$ on the distance $d$ is shown in Fig.
\ref{fig2}d for three different thicknesses of the SiO$_2$ layer on
Si. All the curves are presented for $T_s=600$ K. At the maximum the
frequency of plasmon (\ref{plasma}) matches the resonance in the
emitter $\hbar\omega_p(q)\approx 0.195$ eV at $q\approx 1/2d$.
Deviation from this relation in any direction will result in decrease
of the scaled power. Graphene on a thin silicon oxide film
corresponds to some intermediate situation between pure Si and pure
SiO$_2$.

\begin{figure*}[ptb]
\begin{center}
\includegraphics[width=120mm]{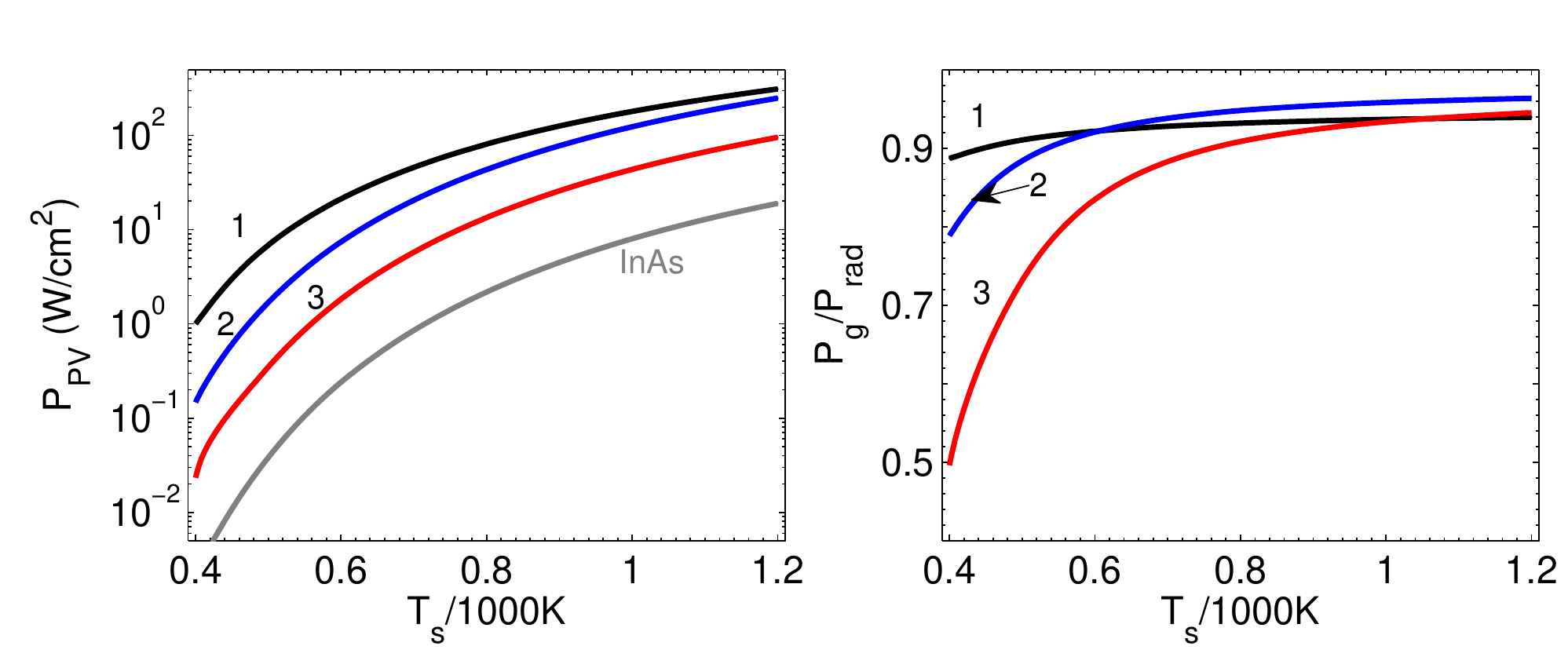}
\vspace{-0.5cm} \caption{(Color online). (left) The output power at
$d=10$ nm as a function of temperature for different barrier heights.
The curves 1, 2 , and 3 corresponds to the barriers 0.18, 0.28, and
0.33 eV, respectively. The emitter resonance frequency is 0.02 eV
higher in all cases. The gray curve shows the result for InAs PV
cell. (right) Fraction of the incident radiation absorbed in the
graphene layer for different barrier heights.} \label{fig3}
\vspace{-0.9cm}
\end{center}
\end{figure*}

For low barrier height the surface electric field at high $T_s$
becomes too small for the carriers separation (the same problem
exists for low bandgap semiconductors). It is therefore interesting
to analyze the behavior of the g/Si cell for different barrier
heights. To perform this analysis we model the emitter with the
dielectric function similar to (\ref{eps_hBN}) but with the
parameters $\varepsilon_{\infty}=5,\ (\omega_L/\omega_T)^2-1=0.5,\
\Gamma/\omega_T=5\times 10^{-3}$, which are close to those for hBN.
The resonance frequency of the reflection coefficient
\begin{equation}\label{res_freq}
    \omega_r=\omega_T(\varepsilon_{em}(0)+1)^{1/2}
    (\varepsilon_{\infty}+1)^{-1/2}
\end{equation}
is the varied parameter, where $\varepsilon_{em}(0)$ is the static
permittivity of the emitter. It is assumed that the barrier height is
somewhat smaller than $\omega_r$. Some results are shown in Fig.
\ref{fig3}. The left panel shows that the output power is smaller for
higher barrier at low $T_s$ but becomes comparable or even larger at
high $T_s$ as the curves 1 and 2 demonstrate. However, when the
barrier is too high $P_{PV}$ becomes smaller at all temperatures. The
reason is that the value of the momentum $q$ in (\ref{plasma}) for
which $\hbar\omega_p > \Phi_b$ becomes significantly larger than
$1/2d$. It reduces the radiative power due to the factor $e^{-2qd}$
and results in the decrease of $P_{PV}$. The output power produced by
InAs PV cell with the bandgap 0.36 eV is shown for comparison in the
same panel. The right panel shows the fraction of the radiative power
absorbed in graphene for different barrier heights. This fraction is
always large especially at high $T_s$. It is somewhat smaller at low
temperatures and high barriers.

Dependence on the barrier height demonstrates that g/Si PV cell is
able to support plasmons in the mid IR range where it is superior
over the semiconductor PV cells. At shorter wavelengths $\lambda <
3.5$ $\mu$m the radiative heat exchange is significantly reduced and
the g/Si cell loses its advantages.

\section{Discussion}

Up to now we assumed that the photodetector is perfect. The typical
responsivity of low bandgap semiconductors is 1 A/W. It is still
larger than the reported value 0.13 A/W \cite{Wan13} for the g/Si
Schottky photodiode. However, the main problem of graphene-based
photodetectors is the coupling of light with graphene. For incoming
propagating waves absorption in graphene is small $\sim\alpha$. To
increase the responsivity one has to couple light with sophisticated
optical structures (see \cite{Wan13} and references therein). These
structures inevitably add losses reducing the effective responsivity.
Additionally, the graphene-Si junction was not optimized. The
configuration used in our scheme does not suffer from the coupling
problem because the heat transfer is realized via the evanescent
waves, for which we have found that 90\% of incoming radiation is
absorbed in a single graphene layer. Therefore, the responsivity for
our configuration can be increased well above 0.13 A/W.

At this moment control of the gap in a few tens of nanometers between
parallel plates is a very challenging problem. However, progress in
this direction is fast. The distances 30-60 nm  are already explored
for RHT experimentally \cite{Zwo12,She12,Shi13} in the sphere-plate
configuration. In the parallel plates configuration the heat transfer
was investigated at distances up to 1 $\mu$m \cite{Ott11,Kra12}.
Large parallel plates separated by a distance of 100 nm or smaller is
also an important problem for many micromechanical applications and
it is actively investigated \cite{Sye13,Woo13}. For this reason we
hope that our proposition can be reality in the near future.

\section{Conclusions}

We considered the near-field TPV element having as a PV cell
graphene-on-silicon Schottky photodiode. Due to presence of graphene
this PV cell has resonant heat exchange with the emitter; a single
layer of graphene absorbs 90\% of incoming radiation that can be
efficiently transformed to photocurrent. Already for
quasi-monochromatic emitter the device demonstrates well advanced
characteristics.




\begin{thebibliography}{99}

\bibitem{Cou01} T. J. Coutts, An overview of thermophotovoltaic
    generation of electricity, Sol. Energy Mater. Sol. Cells {\bf 66},
    443 (2001).

\bibitem{Len14} A. Lenert, D. M. Bierman, Y. Nam, W.
    R. Chan, I. Celanovic, M. Solja\v{c}i\'{c}, and E. N. Wang, A
    nanophotonic solar thermophotovoltaic device, Nature Nano. \textbf{9}, 126
    (2014).

\bibitem{Bas09} S. Basu, Z. M. Zhang and C. J. Fu, Review of
    near-field thermal radiation and its application to energy
    conversion, Int. J. Energy Res. {\bf 33}, 1203 (2009).

\bibitem{Bos12} M. Bosi, C. Ferrari, M. Franceso, M. Pinelli, P. R.
    Spina, and M. Venturini, Thermophotovoltaic generation: A state of the
    art review, in {\it Proceedings of the International
    Conference on Efficiency, Cost, Optimization, Simulation and
    Envinromental Impact of Energy Systems}, 2012, edited by U. Desideri,
    G. Manfrida, E. Sciubba (Firenze University, Perugia, 2012), p. 258.

\bibitem{Pol71} D. Polder and M. Van Hove, Theory of radiative heat
    transfer between closely spaced bodies, Phys. Rev. B {\bf 4},
    3303 (1971).

\bibitem{Pen99} J. B. Pendry, Radiative exchange of heat between
    nanostructures, J. Phys.: Condens. Matter {\bf 11},
    6621 (1999).

\bibitem{Par02} C. H. Park, H. A. Haus, and M. S. Weinberg,
    Proximity-enhanced thermal radiation, J. Phys.
    D {\bf 35}, 2857 (2002).

\bibitem{Mul02} J.-P. Mulet, K. Joulain, R. Carminati, and J.-J.
    Greffet, Enhanced radiative heat transfer at nanometric distances,
    {\bf 6}, 209 (2002).

\bibitem{Har69} C. Hargreaves, Anomalous radiative transfer between
    closely-spaced bodies, Phys. Lett. A {\bf 30}, 491 (1969).

\bibitem{Kit05} A. Kittel, W. M\"{u}ller-Hirsch, J. Parisi, S.-A.
    Biehs, D. Reddig, and M. Holthaus, Near-field heat transfer in a scanning
    thermal microscope, Phys. Rev. Lett. \textbf{95},
    224301 (2005).

\bibitem{Hu_08} L. Hu, A. Narayanaswamy, X. Chen, and G. Chen,
    Near-field thermal radiation between two closely spaced glass
    plates exceeding Planck's blackbody radiation law, Appl.
    Phys. Lett. \textbf{92}, 133106 (2008).

\bibitem{Nar08} A. Narayanaswamy, S. Shen, and G. Chen, Near-field
    radiative heat transfer between a sphere and a substrate, Phys.
   Rev. B {\bf 78}, 115303 (2008).

\bibitem{She09} S. Shen, A. Narayanaswamy, and G. Chen, Surface
    phonon polaritons mediated energy transfer between nanoscale
    gaps, Nano Lett. {\bf 9}, 2909 (2009).

\bibitem{Rou09} E. Rousseau, A. Siria, G. Jourdan, F. Comin, J.
    Chevrier, and J.-J. Greffet, Radiative heat transfer at the nanoscale ,
    Nature. Photon. {\bf 3}, 514 (2009).

\bibitem{Lar06} M. Laroche, R. Carminati, and J.-J. Greffet,
    Near-field thermophotovoltaic energy conversion, J.
    Appl. Phys. {\bf 100}, 063704 (2006).

\bibitem{Jou05} K. Joulain, J.-P.Mulet, F.Marquier, R. Carminati, and
    J.-J. Greffet, Surface electromagnetic waves thermally excited:
    Radiative heat transfer, coherence properties and Casimir forces
    revisited in the near field, Surf. Sci. Rep. {\bf 57}, 59 (2005).

\bibitem{Vol07} A. I. Volokitin and B. N. J. Persson, Near-field
    radiative heat transfer and noncontact friction, Rev. Mod.
    Phys. {\bf 79}, 1291 (2007).

\bibitem{Sve12} V. B. Svetovoy, P. J. van Zwol, and J. Chevrier,
    Plasmon enhanced near-field radiative heat transfer for graphene
    covered dielectrics, Phys. Rev. B {\bf 85}, 155418 (2012).

\bibitem{Jab09} M. Jablan, H. Buljan, and M.
    Solja\v{c}i\'{c}, Plasmonics in graphene at infrared frequencies,
    Phys. Rev. B {\bf 80}, 245435 (2009).

\bibitem{Ili12b} O. Ilic, M. Jablan, J. D. Joannopoulos, I.
    Celanovi\'{c}, H. Buljan and M. Solja\v{c}i\'{c}, Near-field
    thermal radiation transfer controlled by plasmons in graphene,
    Phys. Rev. B {\bf 85}, 155422 (2012).

\bibitem{Zwo12} P. J. van Zwol, S. Thiele, C. Berger, W. A. de Heer,
    and J. Chevrier, Nanoscale radiative heat flow due to surface
    plasmons in graphene and doped silicon, Phys. Rev. Lett.
    {\bf 109} 264301 (2012).

\bibitem{Li_10} X. Li, H. Zhu, K. Wang, A. Cao, J. Wei, C. Li, Y.
    Jia, Z. Li, X. Li, and D. Wu , Graphene-on-silicon Schottky
    junction solar cells, Adv. Mater. {\bf 22}, 2743 (2010).

\bibitem{Ye_12} Y. Ye and L. Dai, Graphene-based Schottky junction
    solar cells, J. Mater. Chem. {\bf 22}, 24224
    (2012).

\bibitem{Mia12} X. Miao, S. Tongay, M. K.
    Petterson, K. Berke, A. G. Rinzler, B. R. Appleton, and A. F. Hebard,
    High efficiency graphene solar cells by chemical doping, Nano Lett.
    {\bf 12}, 2745 (2012).

\bibitem{Wan13} X. Wang, Z. Cheng, K. Xu, H.-K. Tsang, and J.-B. Xu,
    High-responsivity graphene/silicon-heterostructure waveguide
    photodetectors, Nature Photon. \textbf{7}, 888 (2013).

\bibitem{Ili12a} O. Ilic, M. Jablan, J. D. Joannopoulos, I.
    Celanovic and M. Solja\v{c}i\'{c}, Overcoming the black body limit in
    plasmonic and graphene near-field thermophotovoltaic systems, Opt. Express
    {\bf 20}, A366 (2012).

\bibitem{Mes13} R. Messina and P. Ben-Abdallah, Graphene-based
    photovoltaic cells for near-field thermal energy conversion, Sci.
    Rep. \textbf{3}, 1383 (2013).

\bibitem{Yan12b} R. Yan, Q. Zhang, W. Li, I. Calizo, T.
    Shen, C. A. Richter, A. R. Hight-Walker, X. Liang, A.
    Seabaugh, D. Jena, H. G. Xing, D. J. Gundlach, and
    N. V. Nguyen, Determination of graphene work
    function and graphene-insulator-semiconductor band alignment by
    internal photoemission spectroscopy, Appl. Phys. Lett.
    {\bf 101}, 022105 (2012).

\bibitem{Jeo10} H. K. Jeong, K.-J. Kim, S. M. Kim and Y. H. Lee,
    Modification of the electronic structures of graphene by viologen
    Original, Chem. Phys. Lett. \textbf{498}, 168 (2010).

\bibitem{Yi_11} Y. Yi, W. M. Choi, Y. H. Kim, J. W. Kim and S. J.
    Kang, Effective work function lowering of multilayer graphene films
    by subnanometer thick AlO$_x$ overlayers, Appl. Phys. Lett. \textbf{98},
    013505 (2011).

\bibitem{Shi10} Y. Shi, K. K. Kim, A. Reina, M. Hofmann, L.-J. Li and
    J. Kong, Work function engineering of graphene electrode via chemical
    doping, ACS Nano \textbf{4}, 2689 (2010).

\bibitem{Gom09} G. G\'{o}mez-Santos, Thermal van der Waals
    interaction between graphene layers, Phys. Rev. B \textbf{80},
    245424 (2009).

\bibitem{Sve11} V. B. Svetovoy, Z. Moktadir, M. C. Elwenspoek, and
    H. Mizuta, Tailoring the thermal Casimir force with graphene,
    Europhys. Lett. \textbf{96}, 14006 (2011).

\bibitem{Wur82} P. W\"{u}rfel, The chemical potential of radiation,
    J. Phys. C \textbf{15}, 3967 (1982).

\bibitem{Luq03} A. Luque and A. Mart\'{\i}, in \textit{Handbook of
    Photovoltaic Science and Engineering}. Edited by A. Luque and S.
    Hegedus  (John Wiley \& Sons, 2003).

\bibitem{Nar03} A. Narayanaswamy and G. Chen, Surface modes for near
    field thermophotovoltaics, Appl. Phys. Lett.
    \textbf{82}, 3544 (2003).

\bibitem{Par08} K. Park, S. Basu, W.P. King, and Z.M. Zhang,
    Performance analysis of near-field thermophotovoltaic devices
    considering absorption distribution, J.
    Quant. Spectrosc. Radiat. Transfer \textbf{109}, 305 (2008).

\bibitem{Fra11} M. Francoeur, R. Vaillon, and M. P. Meng\"{u}\c{c},
    Thermal impacts on the performance of nanoscale-gap
    thermophotovoltaic power generators,
    IEEE Trans. Energy Convers. \textbf{26}, 686 (2011).

\bibitem{She12} S. Shen, A. Mavrokefalos, P. Sambegoro, and G. Chen,
    Nanoscale thermal radiation between two gold surfaces,
    Appl. Phys. Lett. \textbf{100}, 233114 (2012).

\bibitem{Shi13} J. Shi, P. Li, B. Liu, and S. Shen, Tuning near field
    radiation by doped silicon, Appl. Phys. Lett.
    \textbf{102}, 183114 (2013).

\bibitem{Ott11} R. S. Ottens, V. Quetschke, S. Wise, A. A. Alemi,
    R. Lundock, G. Mueller, D. H. Reitze, D. B. Tanner, and B. F.
    Whiting, Near-field radiative heat transfer between macroscopic
    planar surfaces, Phys. Rev. Lett. \textbf{107}, 014301 (2011).

\bibitem{Kra12} T. Kralik, P. Hanzelka, M. Zobac, V. Musilova, T.
    Fort, and M. Horak, Strong near-field enhancement of radiative
    heat transfer between metallic surfaces, Phys. Rev. Lett. \textbf{109}
    224302 (2012).

\bibitem{Sye13} M. B. Syed Nawazuddin, Micromachined parallel plate
    structures for Casimir force measurement and optical
    modulation, Ph.D. thesis, University of Twente, 2013.

\bibitem{Woo13} D. N. Woolf, Near-field optical forces: photonics,
    plasmonics and the Casimir effect, Ph.D. thesis, Harvard
    University, 2013.

\end{thebibliography}
\end{document}